\documentclass[pre,aps,showpacs]{revtex4}
\usepackage{epsfig}
\usepackage{bm}

\begin{document}

\title{Beyond scaling and locality in turbulence}

\author{\small  A. Bershadskii}
\affiliation{\small {\it ICAR, P.O.\ Box 31155, Jerusalem 91000, Israel}}

\begin{abstract}
An analytic perturbation theory is suggested in order to find
finite-size corrections to the scaling power laws. In the frame of
this theory it is shown that the first order finite-size
correction to the scaling power laws has following form $S(r)
\cong cr^{\alpha_0}[\ln(r/\eta)]^{\alpha_1}$, where $\eta$ is a
finite-size scale (in particular for turbulence, it can be the
Kolmogorov dissipation scale). Using data of laboratory
experiments and numerical simulations it is shown shown that a
degenerate case with $\alpha_0 =0$ can describe turbulence
statistics in the near-dissipation range $r > \eta$, where the
ordinary (power-law) scaling does not apply. For moderate Reynolds
numbers the degenerate scaling range covers almost the entire
range of scales of velocity structure
functions (the log-corrections apply to finite Reynolds number). 
Interplay between local and non-local regimes has been
considered as a possible hydrodynamic mechanism providing the
basis for the degenerate scaling of structure functions and
extended self-similarity. These results have been also expanded on passive 
scalar mixing in turbulence. Overlapping phenomenon between local and 
non-local regimes and a relation between position of maximum of the 
generalized energy input rate and the actual crossover scale between these 
regimes are briefly discussed. 
\end{abstract}


\maketitle

\section{Introduction}
Scaling and related power laws are widely used in physics. In real
situations, however, scaling holds only approximately. As a
consequence, the corresponding scaling power laws hold
approximately as well. On the other hand, discovery of the
extended self-similarity in turbulence \cite{ben1},\cite{ben2} 
and in the critical phenomena (see for a review \cite{sorn},\cite{mps})
shows that universal laws similar to the scaling ones can go far beyond the 
scaling itself, both for velocity \cite{ben1},\cite{ben2} and for different fields 
convected by turbulence \cite{astkb},\cite{bs}.  
The question is: does the extended self-similarity (ESS) can be theoretically 
explored in the frames of general scaling ideas or one needs in a completely new
frames to explain the ESS? In turbulence this problem is related 
to the problem of the so-called near-dissipation range. Theoretical approach to
description of the scaling (the so-called inertial) range of
scales in turbulence requires that the space scales $r \gg \eta$,
where $\eta$ is (Kolmogorov) dissipation or molecular viscous
scale \cite{sa,my}. On the other hand, the near-dissipation range
of scales, for which $r > \eta$ has a more complex, presumably
non-scaling dynamics (see, for instance,
Refs.\cite{nelkin}-\cite{sb2}
and the references cited there). Indeed, for moderate Reynolds
numbers, the near-dissipation range can span most of the available
range of scales \cite{sb}. Different perturbation theories, which
start from scaling as a leading mode, can be developed.
Effectiveness of such theories is usually checked by comparison
with experiments. In the present paper we will explore a generic
analytic expansion of scaling in direction of the dissipation
scale $\eta$. Effectiveness of the suggested perturbation theory
has been shown for the turbulent flows with moderate Reynolds
numbers. Interplay between local and non-local regimes has been
considered as a possible hydrodynamic mechanism providing the
basis for the degenerate scaling of structure functions and
extended self-similarity in turbulent flows.   

\section{Analytic perturbations to scaling}
Let us consider a dimensional function $S(r)$ of a dimensional
argument $r$. And let us construct a {\it dimensionless} function
of the same argument
$$
\alpha (r) = \frac{S^{-1}dS}{r^{-1}dr}.  \eqno{(1)}
$$
If for $L \gg r \gg \eta$ we have no relevant fixed scale (scaling
situation), then for these values of $r$ the function $\alpha (r)$
must be independent on $r$, i.e. $\alpha (r) \simeq const$ for $L
\gg r \gg \eta$. Solution of equation (1) with constant $\alpha$
can be readily found as
$$
S(r) \cong c r^{\alpha}    \eqno{(2)}
$$
where $c$ is a dimension constant. This is the well-known power
law corresponding to the scaling situations.

Let us now consider an analytic theory, which allows us to find generic corrections
of all orders to the approximate power law, related to the fixed small scale $\eta$.
In the non-scaling situation let us denote
$$
f \equiv \ln (S/A), ~~~~~~~~ x \equiv \ln (r/\eta)  \eqno{(3)}
$$
where $A$ and $\eta$ are dimensional constants used for
normalization.

In these variables, equation (1) can be rewritten as
$$
\frac{df}{dx}=\alpha (x) \eqno{(4)}.
$$
In the non-scaling situation $x$ is a dimensionless variable,
hence the dimensionless function $\alpha (x)$ can be non-constant.
Since the 'pure' scaling corresponds to $x \gg 1$ we will use an
analytic expansion in power series
$$
\alpha (x) = \alpha_{0} + \frac{\alpha_1}{x}+...
+ \frac{\alpha_n}{n!x^n}+...  \eqno{(5)}
$$
where $\alpha_n$ are dimensionless constants. After substitution
of the analytic expansion (5) into Eq.\ (4) the zeroth order
approximation gives the power law (2) with $\alpha \cong
\alpha_0$. First order analytic approximation, when one takes only
the two first terms in the analytic expansion (5), gives
$$
S(r) \cong c r^{\alpha_0} [\ln (r/\eta)]^{\alpha_1}.  \eqno{(6)}
$$
This is a generic analytic first order approximation to the
scaling power laws provided by the perturbation theory suggested
above.
\begin{figure} \vspace{-0.7cm}\centering
\epsfig{width=.55\textwidth,file=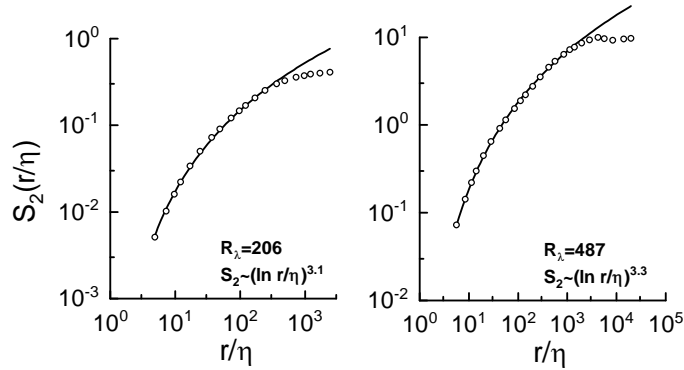} \vspace{-7.5cm}
\caption{The second order structure function $S_2(r/\eta)$ against
$r/\eta$. The experimental data ($R_{\lambda}=206,~487$) \cite{pkw} are shown as
circles. The solid curves are the best fit of (7) to the data,
corresponding to the degenerated scaling. }
\end{figure}
\begin{figure} \vspace{-0.4cm}\centering
\epsfig{width=.55\textwidth,file=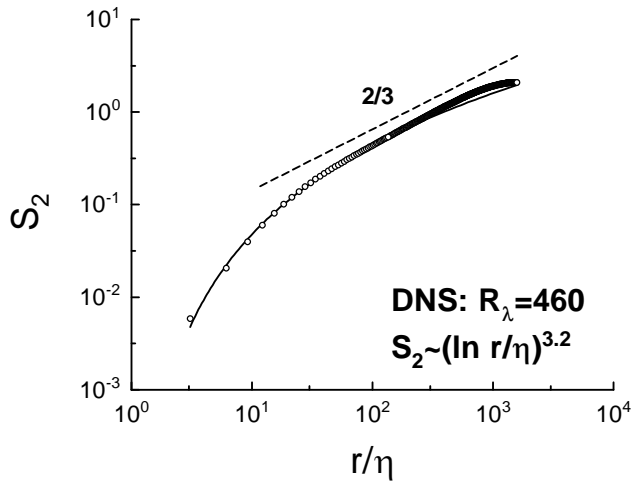} \vspace{-5.5cm}
\caption{The second order structure function $S_2(r/\eta)$ against
$r/\eta$. The DNS data \cite{gfn} ($R_{\lambda}=460$) are shown as
circles. The solid curve is the best fit of (7) corresponding to
the degenerated scaling.  }
\end{figure}

 Corrections of the higher
orders can be readily found in this perturbation theory.

For degenerate case with $\alpha_0=0$ the second term in the expansion (5)
becomes the leading term
$$
S(r) \sim [\ln (r/\eta)]^{\alpha_1}  \eqno{(7)}
$$
 It was already mentioned that there are many different ways 
to develop perturbation theory to scaling. The idea that just 
$\ln (r/\eta)$ is an appropriate parameter for the finite-size corrections to scaling 
in turbulence was suggested in \cite{sb2}. On a physical level this was related in 
\cite{sb2} to the instabilities of the thin vortex tubes (or filaments), which are the 
most prominent hydrodynamical elements of turbulent flows. These instabilities usually 
appear as kinks readily transforming into wave packets propagating along the filaments. 
One can consider such wave packet with scale $r$ and make use of the ``localized induction" 
approximation to the Biot-Savart formula \cite{batc}. In this approximation, contributions 
of portions at distances greater than $\sim r \gg \eta$ to the velocity
fluctuation in the immediate vicinity of any given point on the filament are neglected.
The vortex core radius, which usually associated with $\eta$, provides a cutoff from below.
The vortex filament dynamics in this approximation is described by the equation
$$
\frac{d{\bf X}}{dt}= \frac{\Gamma}{4\pi} \left\{ \ln  \left(
\frac{r}{\eta} \right) \right\} \gamma {\bf b}
$$
where ${\bf X}$ is the position vector of a point on the filament, $\Gamma$ is the vortex strength, 
$\gamma$ is the local curvature, and ${\bf b}$ is the unit
binormal vector of the filament. One can see that in this approximation 
the dependence on $r$ is completely determined by the logarithmic
term $\ln (r/\eta)$ in the dynamic equation. This pure hydrodynamic consideration was 
suggested in \cite{sb2} as a basis for using just $\ln (r/\eta)$ in finite-size 
corrections to scaling in turbulence (see also Section IV below).

\section{Turbulence}
The longitudinal structure function of order $p$ for the velocity field \cite{my}
$$
S_p(r) =\langle |\Delta  u|^p \rangle  \eqno {(8)}
$$
($\Delta u = ({\bf u} ({\bf r} +{\bf x}) - {\bf u} ({\bf r})) \cdot {\bf r}/r$)
calculated for $p=2$, using the data obtained in a wind tunnel at
$R_{\lambda} = 206$ and 487 \cite{pkw}, is shown in Fig.\ 1.
Here $R_\lambda $ is the so-called Taylor microscale Reynolds
number. The experiment with a flow, which was a combination of the wake
and homogeneous turbulence behind a grid, is described in Ref.\
\cite{pkw}. We invoke Taylor's hypothesis \cite{my} to equate temporal
statistics to spatial statistics.

\begin{figure} \vspace{0cm}\centering
\epsfig{width=.55\textwidth,file=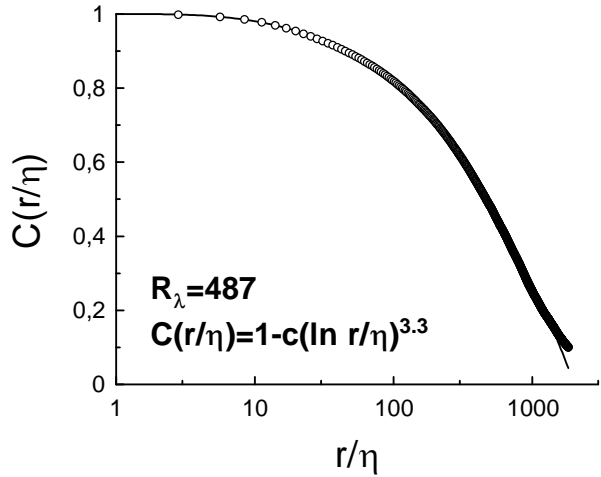} \vspace{-5.5cm}
\caption{Correlation function for velocity fluctuations against
$\log r/\eta$ for $R_{\lambda} = 487$ (circles) \cite{pkw}. The solid curve is
drawn in the figure to indicate agreement with the degenerate
scaling shown in Fig.\ 1.}
\end{figure}

The solid curves in these figures are the best fit by equation (7)
corresponding to the degenerate scaling. Figure  2 shows $S_2
(r)$ calculated using data from a high-resolution direct numerical
simulation of homogeneous steady three-dimensional turbulence
\cite{gfn}, corresponding to $1024^3$ grid points and
$R_{\lambda}=460$. The solid curve in this figure also corresponds
to the best fit by equation (7). As far as we know Eq.\ (7) was
suggested as an empirical approximation of the structure functions
for the first time in our earlier paper \cite{sb}. Application of
these results to correlation functions \cite{my} is in good
agreement with the structure functions analysis (cf. for instance,
Figs. 1 and  3).

Figure 4 shows the structure functions of different orders $p$ for
the wind-tunnel data ($R_{\lambda} = 487)$ and the degenerate
scaling (7) is shown in the figure as the solid curves. One can
see that the degenerate scaling requires two fitting constants
just as the ordinary scaling does, and, in the examples discussed
above, the range of scales covered by the degenerate scaling is
about two decades.

\section{A possible hydrodynamic scenario}

In order to understand a possible hydrodynamic mechanism of the
phenomenon described above let us recall that in isotropic
turbulence a complete separation of local and non-local interactions is 
possible in principle. It
was shown by Kadomtsev \cite{kad} that this separation plays a
crucial role for the local Kolmogorov's cascade regime with scaling energy spectrum
$$
E(k) \sim \bar{\varepsilon}^{2/3} k^{-5/3}  \eqno{(9)}
$$
where $\bar{\varepsilon}$ is the average of the
energy dissipation rate, $\varepsilon$, $k \sim 1/r$ is the wave-number. This
separation should be effective for the both ends. That is, if there
exists a solution with the local scaling (9) as an asymptote, then there should also
exist a solution with the non-local scaling asymptote. Of course, the two solutions with
these asymptotes should be alternatively stable (unstable) in different regions
of scales. It is expected, that the local (Kolmogorov's) solution is stable (i.e.
statistically dominating) in inertial range (that means instability of the non-local
solution in this range of scales).

\begin{figure} \centering \vspace{-0.3cm}
\epsfig{width=.8\textwidth,file=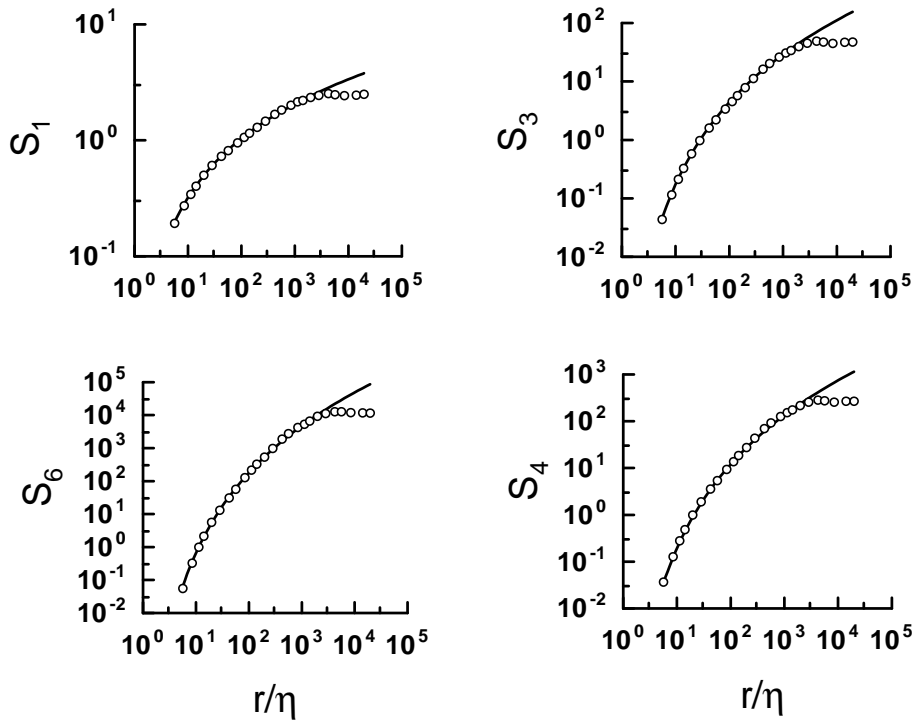} \vspace{-9cm}
\caption{As in Fig.\ 1 ($R_{\lambda}=487$) but for different orders of the structure
function.}
\end{figure}
Roughly speaking, in non-local solution for small scales $r$ only
non-local interactions with large scales $L$ ($1 \gg r/L$) are
dynamically significant and the non-local interactions is
determined by large scale strain/shear. This means that one should
add to the energy flux $\bar{\varepsilon}$-parameter (which is a
governing parameter for the both solutions) an additional
parameter such as the strain $s$ for the non-local solution. As
far as we know it was noted for the first time by Nazarenko and
Laval \cite{nl} that dimensional considerations applied to
the non-local asymptotic result in the power-law energy spectrum
$$
E(k) \sim \frac{\bar{\varepsilon}}{s}~ k^{-1}  \eqno{(10)}
$$
both for two- and three-dimensional cases. Linear dependence of
the spectrum (10) on $\bar{\varepsilon}$ is determined by the linear
nature of equations corresponding to the non-local asymptotic that
together with the dimensional considerations results in (10)
\cite{nl}. Interesting numerical simulations were performed in \cite{ldn}. 
In these simulations local and non-local interactions have been alternatively 
removed. For the first case a tendency toward a spectrum flatter than '-5/3' is observed 
near and beyond the separating scale  (beyond which local interactions are ignored), 
that supports Eq. (10) (see also below).

Following to the perturbation theory suggested in Section II both
local and non-local regimes can be corrected. The first order correction is
$$
E(k) \sim k^{-5/3} [\ln (k_d/k)]^{\gamma}  \eqno{(11)}
$$
and
$$
E(k) \sim k^{-1} [\ln (k_d/k)]^{\beta}  \eqno{(12)}
$$
(where $k_d\equiv1/\eta$) for the local and non-local 
regimes respectively ($\gamma$ and
$\beta$ are dimensionless constants). In the case when $\beta
\gg \gamma$ the fist order finite-size correction in the
non-local regime becomes substantial much 'earlier' (i.e. for
smaller $k$) than for the local (Kolmorov's) regime (see Eq.\
(5)). This can result in viscous stabilization of the non-local
regime (cf \cite{ldn}) and, as a consequence, in the so-called 'exchange of
stability' phenomenon at certain $k_c=1/r_c$. That is, for $k <
k_c$ the Kolmogorov's regime is stable and the non-local regime is
unstable, whereas for $k > k_c$ the Kolmogorov's regime is
unstable and the non-local regime is stable. For this scenario, at
$k=k_c$ the Kolmogorov's regime is still asymptotically
scale-invariant (i.e. Eq.\ (9) gives an adequate approximation for
this regime), while for the non-local regime the first order
correction is substantial (i.e. Eq.\ (12) should be used at
$k=k_c$ for the non-local regime). In this scenario the
Kolmogorov's regime plays significant role in the viscous
stabilization of the non-local regime for $k > k_c$ (cf \cite{ldn}), but for these
$k$ the non-local regime becomes statistically dominating instead
of the Kolmogorov's one.
\begin{figure} \vspace{-0cm}\centering
\epsfig{width=.55\textwidth,file=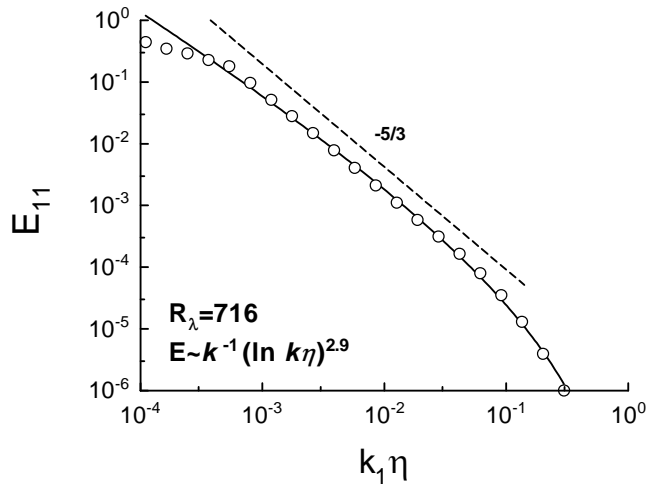} \vspace{-5.5cm}
\caption{Longitudinal energy spectrum against $k\eta$ in log-log scales.
The data (circles) correspond to a nearly isotropic (active-) grid
turbulence $R_{\lambda} = 716$ \cite{kcm}. The solid curve is
drawn in the figure to indicate agreement with the Eq.\ (12),
while the dashed straight line indicates the "-5/3" Kolmogorov's law. }
\end{figure}
Figure 5 shows a spectrum measured in nearly isotropic turbulence
downstream of an active grid at $R_{\lambda} = 716$ (the data are
reported in \cite{kcm}). The solid curve is drawn in the figure to
indicate correspondence of the data to the equation (12)
(non-local regime), while the dashed straight line indicates the
Kolmogorov's "-5/3" law to show possible interplay between the
Kolmogorov's and non-local regimes at this Reynolds number.

If one tries to estimate the second order structure function $S_2 (r)$ as \cite{my}
$$
S_2(r) \sim \int_{1/r}^{1/\eta} E(k) dk   \eqno{(13)}
$$
for $r < r_c$ using Eq.\ (12), one obtains
$$
S_2(r) \sim [\ln (r/\eta)]^{\zeta_2}  \eqno{(14)}
$$
with $\zeta_2 =\beta + 1$ (cf Section 3). 
For instance, for the data shown in Fig. 5 we obtain $\zeta_2 \simeq 3.9$. 
Comparing figures 1,2 and 5 one can see that $\alpha$ is monotonically increased with $Re_{\lambda}$. \\

Let us generalize (13) introducing an effective spectrum $E_n (k)$ 
$$
S_n(r) \sim \int_{1/r}^{1/\eta} E_n(k) dk   \eqno{(15)}
$$ 
($E_2 (k) \equiv E (k)$). Then using the dimensional considerations we obtain for the non-local regime
$$
E_n (k) \sim \left(\frac{\bar{\varepsilon}}{s} \right)^{n/2} ~ k^{-1} \eqno{(16)}
$$
The first order correction is
$$
E_n(k) \sim \left(\frac{\bar{\varepsilon}}{s} \right)^{n/2} ~ k^{-1} [\ln (k_d/k)]^{\beta_n}  \eqno{(17)}
$$
Substituting (17) into (15) one obtains
$$
S_n(r) \sim   \left(\frac{\bar{\varepsilon} }{s} \right)^{n/2} ~[\ln (r/\eta)]^{\zeta_n}  \eqno{(18)}
$$
with $\zeta_n=\beta_n + 1$.

Thus the non-local regime can provide hydrodynamic basis for the degenerate scaling
of the structure functions.

\section{Extended Self-Similarity}

\begin{figure} \vspace{-0cm}\centering
\epsfig{width=.55\textwidth,file=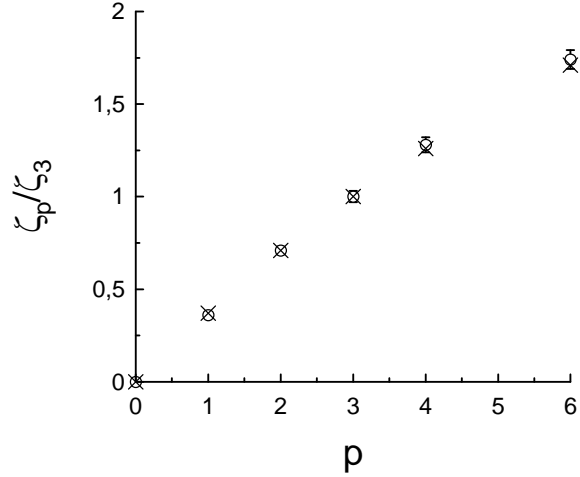} \vspace{-5.5cm}
\caption{Normalized exponents $\zeta_p/\zeta_3$ against $p$ for
$R_{\lambda}=487$ (circles). The exponents were obtained using the
data shown in Figs. 1,4. Crosses are the ESS exponents
obtained for the atmospheric turbulence data at $R_{\lambda}
=10340$ \cite{sd}.}
\end{figure}

In our previous paper \cite{sb} we discussed possible relation
between ESS and structure functions given by Eq. (7). Now we can
elaborate this relation with more details. While in the
near-dissipation range we observe the degenerate scaling an
ordinary scaling (presumably Kolmogorov's one) has been certainly
observed in the inertial range of scales for very large Reynolds
numbers (see, for instance, \cite{my},\cite{sd}). Therefore, one
can expect the existence of a crossover scale $r_c$ from the
ordinary scaling to the degenerate one (see previous section). At
the scale $r=r_c$ we can use a continuity condition between
ordinary scaling (2) and degenerate scaling (7)
$$
cr_c^{\zeta'_2}=A[\ln (r_c/\eta]^{\zeta_2}   \eqno{(19)}
$$
for the structure functions of all orders ($c$ and $A$ are some constants).
Let us denote in general case
$$
S_p(r)=A_p'r^{\zeta_p'};  ~~~~~
S_p(r)=A_p[\ln(r/\eta)]^{\zeta_p} \eqno{(20)}
$$
in inertial and near-dissipation ranges respectively.

Then condition of compatibility of the continuity equations (19)
for different order ($m$ and $n$) structure functions
$$
A_m'r_c^{\zeta_m'} = A_m[\ln(r_c/\eta)]^{\zeta_m};~~~
A_n'r_c^{\zeta_n'} = A_n[\ln(r_c/\eta)]^{\zeta_n}  \eqno{(21)}
$$
can be of two kinds.

The first kind is applicable for arbitrary value of $r_c$ and has
the form
$$
\frac{\zeta_m}{\zeta_n}= \frac{\zeta_m'}{\zeta_n'};~~~~~
\left( \frac{A_m'}{A_m} \right)^{\zeta_n}=\left( \frac{A_n'}{A_n} \right)^{\zeta_m}
   \eqno{(22)}
$$
while another kind is applicable for a fixed value of $r_c$ only.
Below we will be interested in the first kind of the compatibility condition.
This condition, in particular, gives a power-law relation between moments of
different orders
$$
S_m (r) \sim S_n^{\beta_{m,n}} (r),   \eqno{(23)}
$$
with the same exponent
$$
\beta_{m,n} = \frac{\zeta_m'}{\zeta_n'}= \frac{\zeta_m}{\zeta_n}  \eqno{(24)}
$$
for both inertial and near-dissipation ranges (see Eq.\ (22)). The
last phenomenon was previously observed in different turbulent
flows and was called Extended Self-Similarity (ESS)
\cite{ben1},\cite{ben2},\cite{sb}. It is important to note that at ESS not only 
the scalings (23) are the same in both regimes inertial and dissipative, but also the 
prefactors are the same. Figure 6 shows as circles the exponents
$\zeta_p$ obtained from Figs. 1,4 and normalized by the exponent
$\zeta_3$. For comparison we show by crosses in this figure the
normalized exponents obtained using ESS in the atmospheric
turbulence for large $R_{\lambda} = 10340$ \cite{sd}. 

It should be noted that a high-resolution numerical simulation reported in 
a recent paper \cite{sys} for {\it low-Reynolds-number} ($R_{\lambda} \approx 10-60$) 
flows shows no hint of scaling-like behavior of the velocity increments even when ESS is applied. 
That should return us to the Section II and recall us that the degenerate scaling 
in the form of Eq. (7) is only a {\it first} non-trivial term in the perturbation theory. 
Moreover, the question of convergence of the perturbation theory used in Section II become 
significant at the {\it low-Reynolds-number}, when the dissipation scale is effectively 
too close to the scales under consideration. Though, the authors of Ref. \cite{sys} claim that: 
'the DNS scaling exponents of velocity gradients agree well with those deduced, using a recent 
theory of anomalous scaling, from the scaling exponents of the longitudinal 
structure functions at infinitely high Reynolds numbers. This suggests that
the asymptotic state of turbulence is attained for the velocity gradients at far lower
Reynolds numbers than those required for the inertial range to appear'. This circle of problems 
can be also related to the possibility of fluctuations of the dissipation scale \cite{sys},\cite{y}. 
For the cases when these fluctuations are significant a set of the expansions might be needed or 
(alternatively, for moderate Reynolds numbers) calculation should involve averaging over the 
fluctuating cut-off $\eta$. In the last case the experimental and numerical data suggest that 
the average value of $\eta$ should not deviate strongly from the Kolmogorov dissipation scale (at least 
for the moderate values of Reynolds number). 

Though, the problem with the small Reynolds numbers can be deeper. We do not know 
whether the non-local regime appears at the same Reynolds number as the local one. The energy 
balance of the non-local regime is more complex than that of the local one. Actually the non-local 
regime needs in the local one (even statistically unstable) for this balance (cf \cite{ldn}). 
Therefore, {\it critical} Reynolds number for appearance of the (stable) non-local regime can be larger 
than that for the local regime.

\section{Generalized energy input rate}

Figures 2 and 5 show that for sufficiently large Reynolds numbers, providing 
a visible inertial interval, there is an overlapping between the two scalings: 
non-local and local (Kolmogorov). This overlapping is based on the very nature 
of the stability exchange between the two {\it statistical} regimes. Therefore, it 
is not a simple task to determine the scale $r_c$ from the $S_n (r)$-data. 
A more fine information one can infer studying 
$$
D_{LLL} (r) =  \langle \Delta u^3 \rangle  
$$
(cf Eq.(8) and see \cite{my},\cite{sb2},\cite{gfn}). Unlike $S_3$ 
most positive contributions are canceled by negative ones in 
$D_{LLL}$, and only the slight asymmetry of the $\Delta u$ probability density contributes 
to $D_{LLL}$. However, this asymmetry has a fundamental nature. Therefore, one can expect that the above used dimensional considerations 
can be also used for $ D_{LLL}$ for the both non-local and local regimes. For the local 
(Kolmogorov) regime one has \cite{my},\cite{sb2},\cite{gfn} 
$$
-D_{LLL} \sim \bar{\varepsilon}
 ~ r  \eqno(25)
$$ 
while for the degenerate scaling
$$
-D_{LLL} \sim  \ \left(\frac{\bar{\varepsilon}}{s} \right)^{3/2} ~[\ln (r/\eta)]^{\delta}  
\eqno{(26)}
$$
Figure  7 shows $-D_{LLL}$ (as circles) and $-D_{LLL}/\bar{\varepsilon} r$ (as crosses) 
calculated using data from the direct numerical
simulation of homogeneous steady three-dimensional turbulence
\cite{gfn}. The solid curve in this figure corresponds
to the best fit by equation (26). In this figure one can see that the overlapping between 
the two regimes exists even for the $D_{LLL}$ (cf Fig. 2). 
However, a generalized energy input rate related 
to $D_{LLL}$ \cite{sb2} could give a key to our problem. 
\begin{figure} \vspace{-0.4cm}\centering
\epsfig{width=.65\textwidth,file=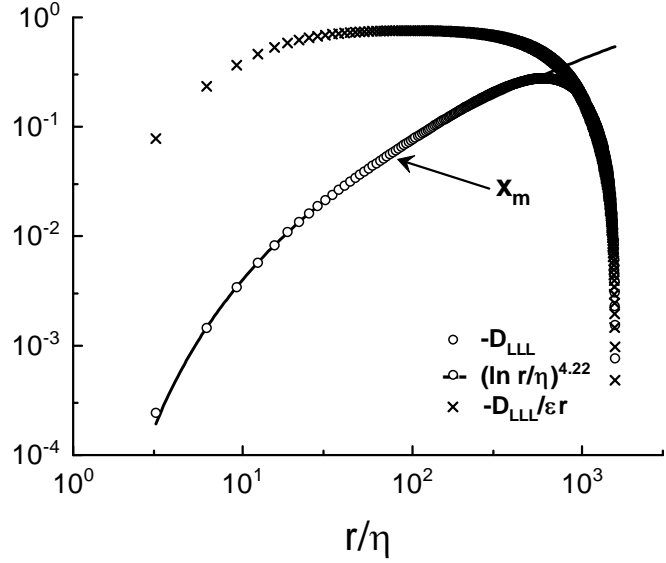} \vspace{-6.5cm}
\caption{ The DNS data \cite{gfn} ($R_{\lambda}=460$) are shown against
$r/\eta$ as the symbols: $-D_{LLL}$-circles and $-D_{LLL}/\bar{\varepsilon} r$-crosses. 
The solid curve is the best fit of (26) corresponding to
the degenerated scaling.  }
\end{figure}
Let us consider the Navier-Stokes equations for a viscous incompressible
fluid, given by
$$
\frac{\partial u_i}{ \partial t} + u_j \frac{\partial u_i}{
\partial x_j} = - \frac{1}{\rho} \frac{\partial p}{ \partial x_i} +
\nu \frac{\partial^2 u_i}{\partial x_j^2} + f_i ({\bf x}, t),
\eqno{(27a)}
$$
$$
\frac{\partial u_i}{\partial x_i} =0, \eqno(27b)
$$
where ${\bf f} ({\bf x}, t)$ is random force,  
$\nu$ is kinematic viscosity and $\rho$ is the fluid
density. $\bf f$ will be assumed to be Gaussian with zero mean 
and a rapidly oscillating character, or a $\delta$-correlation in time.
The second-rank correlation tensor
$$
\langle f_i ({\bf x}+{\bf r}, t+\tau) f_j ({\bf x}, t) \rangle \sim F_{ij}
({\bf r}) \delta (\tau).   \eqno{(28)}
$$
defines such fields. 
\begin{figure} \vspace{-0.4cm}\centering
\epsfig{width=.45\textwidth,file=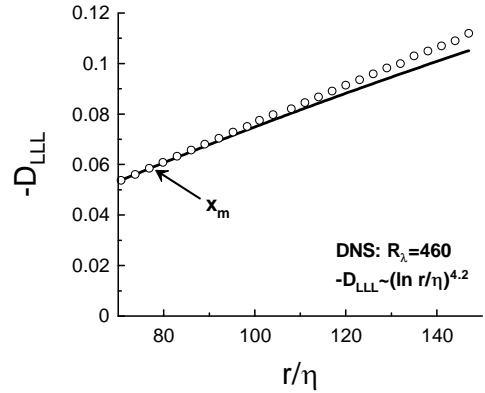} \vspace{-4.5cm}
\caption{$-D_{LLL}$ against
$r/\eta$. The same data as in Fig. 7 but with higher resolution 
in a small vicinity of the $x_m$ scale. 
The solid curve is the best fit of (26) corresponding to
the degenerate scaling.  }
\end{figure}
\begin{figure} \vspace{-0.4cm}\centering
\epsfig{width=.45\textwidth,file=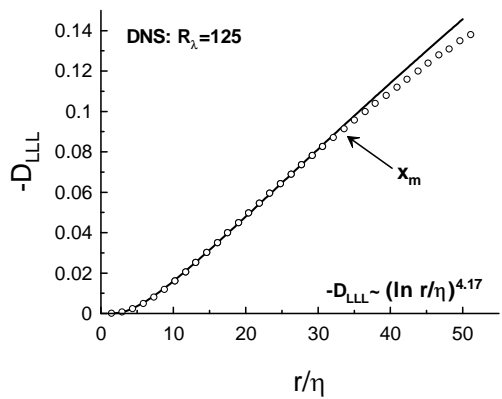} \vspace{-4.5cm}
\caption{ The same as in Fig. 8 but for $R_{\lambda} = 125$.}
\end{figure}
It was shown in \cite{nov} using Eqs. (27) and (28) that $S_2$ and $D_{LLL}$ 
are related by the equation
$$
D_{LLL} = 6\nu \frac{dS_2}{dr} - \frac{2}{r^4} \int_0^r x^4 F_{ii} (x)
dx,  \eqno{(29)}
$$
It was also shown in \cite{nov} that
$$
F_{ii} (0)= 2 \langle \bar{\varepsilon} \rangle. \eqno{(30)}
$$
$F_{ii}$ corresponds to an external energy input rate \cite{nov} 
(cf Novikov's relation \cite{nov}: $\langle
f_i({\bf x},t) v_j ({\bf x}',t)\rangle =\frac{1}{2}~F_{ij} ({\bf
x}-{\bf x}')$).

For $r \ll L$ \cite{nov}
$$
D_{LLL} \simeq 6\nu \frac{dS_2}{dr} - \frac{4}{5} \langle \bar{\varepsilon}
\rangle r. \eqno{(31)}
$$
When the viscous term can be neglected this gives the Kolmogorov's law (25). 

One can formally rewrite (29) as
$$
D_{LLL} =  - \frac{2}{r^4} \int_0^r x^4 \tilde{F} (x) dx, \eqno{(32)}
$$
where the generalized energy input rate is defined as \cite{sb2}
$$
\tilde{F} (x) \equiv F_{ii} (x) -\frac{3\nu}{x^4} \frac{d (x^4
dS_2/dx)}{dx}. \eqno{(33)}
$$  
 
Without the second (viscous) term in the right-hand side of Eq. (33) one obtains 
the Kolmogorov law (25). Therefore, one can associate the first term in the right-hand side of the Eq. (33) with 
the local (Kolmogorov) regime while the second term can be associated with the non-local one. 
Gradients of these terms can be associated with local and non-local interaction strengths respectively. 
Then balance of the local and non-local interaction strengths should be reached at the point 
given by the equation $ d\tilde{F} (x)/dx =0$. Let us denote position of the balance point as $x=x_m$. 
One can expect that the crossover scale $r_c $ 
between the local and non-local regimes (see Section IV) coincides with the balance point of the 
local and non-local interaction strengths $x_m$, 
i.e. $r_c=x_m$.  It was shown in \cite{sb2} that position of the generalized energy input rate maximum 
$x_m \cong r_m/1.22 $, where $r_m$ is position of $-D_{LLL}(r)$ maximum. 
Since $r_m$ can be obtained from the data this gives us a possibility to estimate $r_c$ from the data as well. 
For instance, we know \cite{sb2} that $r_m/\eta \sim R_{\lambda}^{0.7}$, hence
$r_c/ \eta \sim R_{\lambda}^{0.7}$. 
\begin{figure} \vspace{-0.4cm}\centering
\epsfig{width=.65\textwidth,file=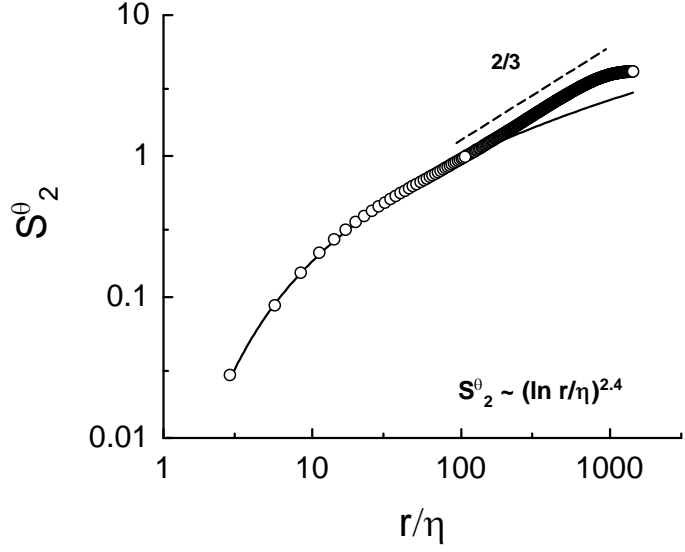} \vspace{-7cm}
\caption{$S^{\theta}_2$ against
$r/\eta$. The DNS data (circles) of homogeneous isotropic turbulence 
described in \cite{wg}, P\'eclet number $P_{\lambda}=427$ and the Schmidt number is unity 
(i.e $P_{\lambda}=R_{\lambda}$). The solid curve is the best
fit to the degenerate scaling (40) (the dashed straight line indicates the Obukhov-Corrsin 
ordinary scaling \cite{my},\cite{wg}).}
\end{figure}
In Fig. 7 we indicated position of the scale $x_m$ by an arrow and one can see 
that overlapping of the two regimes can be substantial even for $D_{LLL}$. In figure 8 
we show the same data as in Fig. 7 
but with a considerably higher resolution in a small vicinity of the $x_m$ scale. The 
degenerate scaling fit (solid curve) indeed seems to begin its declination from the data 
in a small vicinity of the $x_m$-scale in this case. To support this point we show in 
figure 9 analogous situation observed for considerably smaller Reynolds number 
$Re_{\lambda}= 125$ (see also next Section).  \\
 
\section{Passive scalar}
\begin{figure} \vspace{-0.4cm}\centering
\epsfig{width=.7\textwidth,file=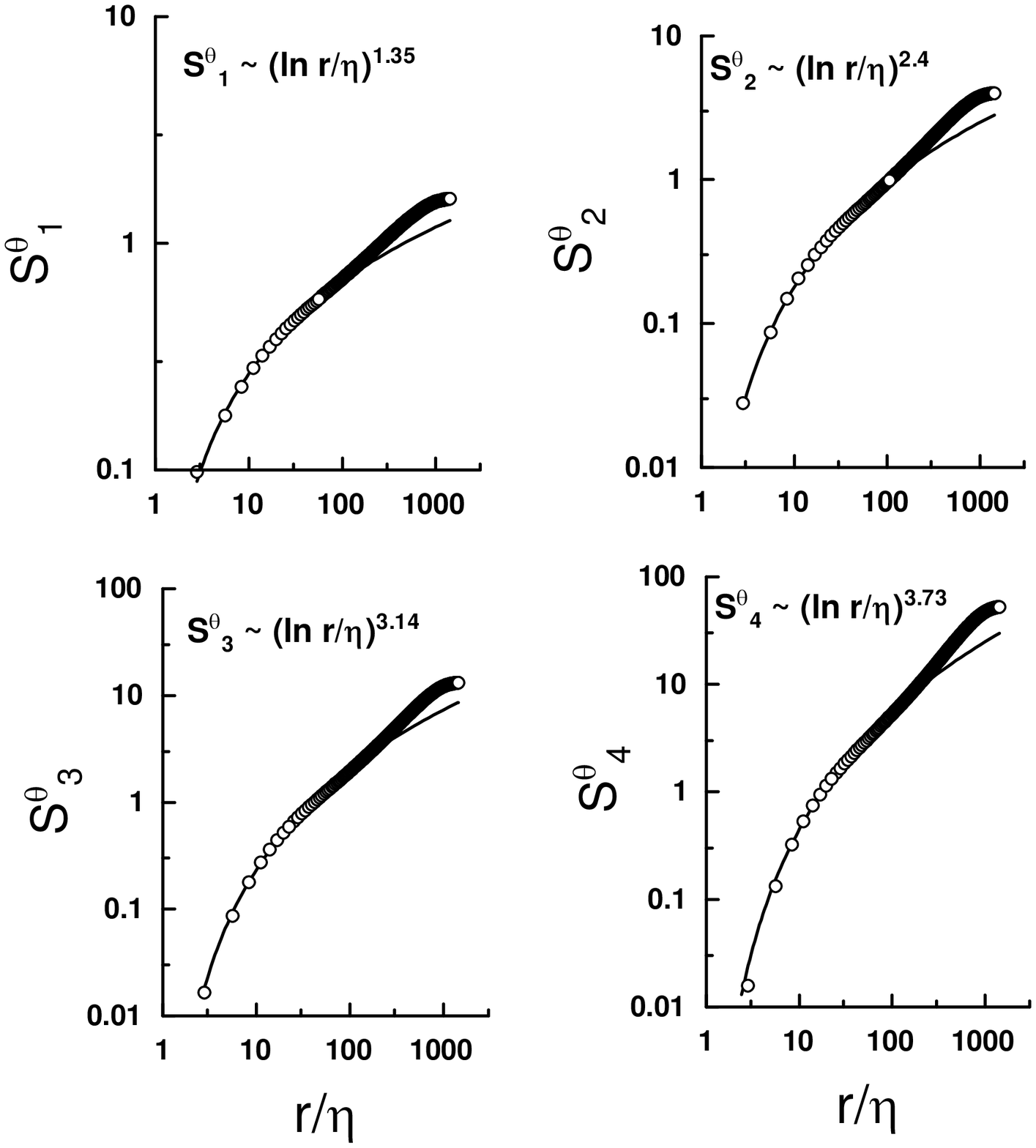} \vspace{-4cm}
\caption{$S^{\theta}_n$ against
$r/\eta$. The DNS data (circles) of homogeneous isotropic turbulence 
described in \cite{wg}, P\'eclet number $P_{\lambda}=427$ and the Schmidt number is unity 
(i.e $P_{\lambda}=R_{\lambda}$). The solid curves are the best
fit to the degenerate scaling (44).}
\end{figure}
The inertial-convective regime of a passive scalar field transport
in isotropic turbulence for large Reynolds numbers is usually
described using Obukhov-Corrsin theory \cite{my}. This theory predicts
scaling spectrum for the passive scalar ($\theta$) fluctuations
with the scaling exponent equal to "-5/3"
$$
E^{\theta} (k) \sim \bar{\varepsilon}^{-1/3} \bar{\varepsilon}_{\theta} 
k^{-5/3} \eqno{(34)}
$$
where $\bar{\varepsilon}_{\theta}$ is the average value of dissipation
rate of scalar variance, ${\varepsilon}_{\theta}$ (cf Eq. (9)). In the experiments,
however, this regime is not usually observed (especially, for
moderate Reynolds numbers and in turbulent shear flows). In the
last case the experiments indicate a strong dependence of the
passive scalar spectra on the Reynolds number (see, for instance,
Refs. \cite{sreeni1}, \cite{warhaft} and references therein). 
In the vein of the Section IV the dimensional considerations applied to
the non-local asymptotic result in the power-law spectrum
$$
E^{\theta}(k) \sim \frac{\bar{\varepsilon}_{\theta}}{s}~ k^{-1}  \eqno{(35)}
$$ 
(cf Eq. (10) and \cite{nl},\cite{ldn},\cite{vid}). Accordingly,  both scaling regimes 
(34) and (35) can be corrected (for simplicity we restrict ourselves by the case when 
the Schmidt number is unity \cite{wg}). 
The first order correction is
$$
E_{\theta} (k) \sim k^{-5/3} [\ln (k_d/k)]^{\gamma}  \eqno{(36)}
$$
and
$$
E^{\theta} (k) \sim k^{-1} [\ln (k_d/k)]^{\beta}  \eqno{(37)}
$$
for the local and non-local regimes respectively ($\gamma$ and $\beta$ are dimensionless 
constants, different from those in Eqs. (11),(12) ). 

One can estimate the second order structure function for the passive scalar fluctuations
$$
S^{\theta}_2 (r) = \langle |\Delta \theta|^2 \rangle = \langle |\theta ({\bf r} +{\bf x}) 
- \theta ({\bf r})|^2 \rangle  \eqno{(38)}
$$ 
as
$$
S^{\theta}_2(r) \sim \int_{1/r}^{1/\eta} E^{\theta}(k) dk   \eqno{(39)}
$$
(cf  Eq. (13)). Then for $r < r_c$ using Eqs.\ (37) and (39) one obtains
$$
S^{\theta}_2(r) \sim [\ln (r/\eta)]^{\zeta_2}  \eqno{(40)}
$$
with $\zeta_2 =\beta + 1$ (cf Section IV). 

Figure 10 shows $S^{\theta}_2$ against $r/\eta$ for the
DNS data of homogeneous isotropic turbulence described in \cite{wg}, 
P\'eclet number $P_{\lambda}=427$ and the Schmidt number is unity 
(i.e. $P_{\lambda}=R_{\lambda}$). The solid curve is the best
fit to the degenerate scaling (40) (the dashed straight line indicates the Obukhov-Corrsin 
ordinary scaling \cite{my},\cite{wg}).
\begin{figure} \vspace{-0cm}\centering
\epsfig{width=.65\textwidth,file=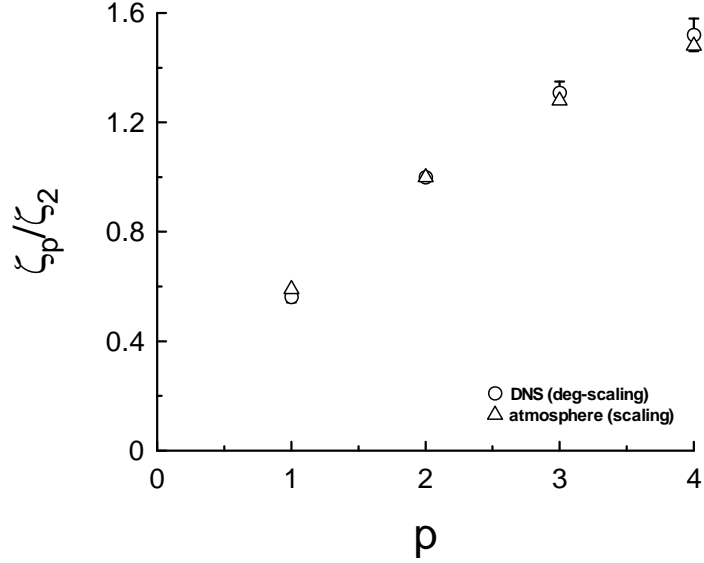} \vspace{-6cm}
\caption{Exponents extracted from Fig. 11 (circles) and for
the fully developed atmospheric turbulence (ordinary scaling in
the inertial interval, triangles \cite{lov}).)}
\end{figure}
As in Section IV we can generalize (39) introducing an effective spectrum $E^{\theta}_n (k)$ 
$$
S^{\theta}_n(r) = \langle |\Delta \theta|^n \rangle\sim \int_{1/r}^{1/\eta} E^{\theta}_n(k) dk   \eqno{(41)}
$$ 
($E^{\theta}_2 (k) \equiv E^{\theta} (k)$). 
Then using the dimensional considerations we obtain for the non-local regime
$$
E^{\theta}_n (k) \sim \left(\frac{\bar{\varepsilon}_{\theta}}{s} \right)^{n/2} ~ k^{-1} \eqno{(42)}
$$
The first order correction is
$$
E^{\theta}_n(k) \sim \left(\frac{\bar{\varepsilon}_{\theta}}{s} \right)^{n/2} ~ k^{-1} [\ln (k_d/k)]^{\beta_n}  \eqno{(43)}
$$
Substituting (43) into (41) one obtains
$$
S^{\theta}_n(r) \sim   \left(\frac{\bar{\varepsilon}_{\theta} }{s} \right)^{n/2} ~
[\ln (r/\eta)]^{\zeta_n}  \eqno{(44)}
$$
with $\zeta_n=\beta_n + 1$.

Figure 11 shows $S^{\theta}_n$ against $r/\eta$ for the
DNS data of homogeneous isotropic turbulence described in \cite{wg}, 
P\'eclet number $P_{\lambda}=427$ and the Schmidt number is unity 
(i.e $P_{\lambda}=R_{\lambda}$). The solid curves are the best
fit to the degenerate scaling (44). \\

One can readily expand the conclusions of the section V regarding 
Extended Self-Similarity on the case of passive scalar 
and we show effectiveness of this in figure 12. The data for degenerate scaling (circles) 
were taken from the degenerate scaling shown in Fig. 11 whereas the data for the ordinary 
scaling (triangles) were taken from a high Reynolds number atmospheric experiment \cite{lov}.\\    
\begin{figure} \vspace{-0.4cm}\centering
\epsfig{width=.65\textwidth,file=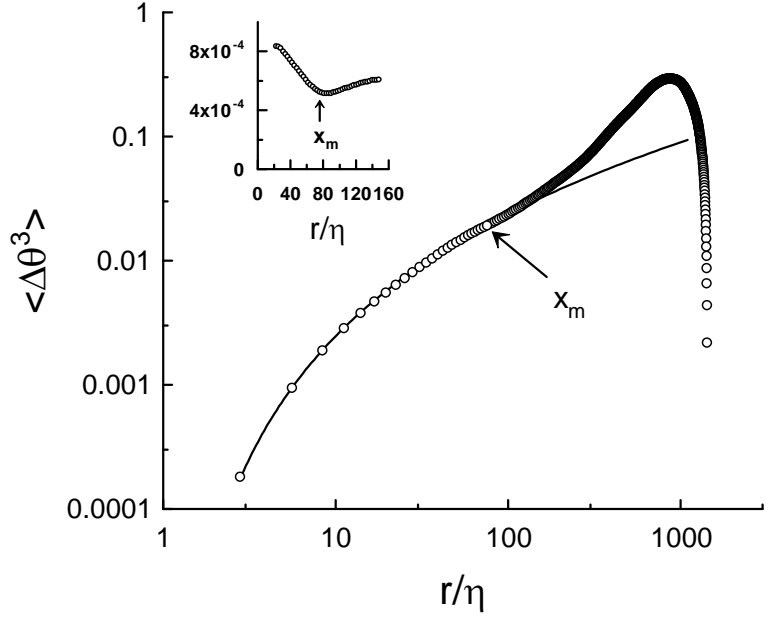} \vspace{-5.5cm}
\caption{$\langle \Delta \theta^3 \rangle $ against
$r/\eta$. The DNS data (circles) of homogeneous isotropic turbulence 
described in \cite{wg}, P\'eclet number $P_{\lambda}=427$ and the Schmidt number is unity 
(i.e $P_{\lambda}=R_{\lambda}$). The solid curve is the best
fit to the degenerate scaling. The inset shows with high resolution 
the derivative of $\langle \Delta \theta^3 \rangle$ in a vicinity of its minimum. 
The arrow indicates position of the generalized energy input 
rate maximum $x_m \simeq r_c$ calculated using $D_{LLL}$.}
\end{figure}
The overlapping phenomenon (Section VI) takes also place for the 
$S^{\theta}_n$ as for $S_n$. In $\langle \Delta \theta^3 \rangle $ like in 
$D_{LLL}$ most positive contributions are canceled by negative ones and only 
the slight asymmetry of the $\Delta \theta$ probability density contributes 
to $\langle \Delta \theta^3 \rangle $. However, similarly to $D_{LLL}$ 
this asymmetry has a fundamental nature. Therefore, one can expect that 
$\langle \Delta \theta^3 \rangle $ also can be used in order to determine 
$r_c$. Moreover, since $\langle \Delta \theta^3 \rangle $ as function of 
$r/\eta$ has a turnover point, as one can see in figure 13, the turnover point 
should indicate the point where the actual declination of the degenerate scaling 
fit from the data should begin. In the turnover point derivative of the function 
$\langle \Delta \theta^3 \rangle $ has its minimum. In the inset to Fig. 13 
we show with high resolution the derivative in a vicinity of its minimum. We also 
indicate with arrow the position of local maximum of the generalized energy input 
rate $x_m \simeq r_c$ calculated using $D_{LLL}$ for this case (cf Section VI).   \\

One can see that the suggested perturbation theory with the
degenerate scaling is an appropriate tool for description of data
in the near-dissipation range. For modest Reynolds numbers the
degenerate scaling applies to nearly the entire range of scales.
Crossover between the degenerate and ordinary scaling provides an
explanation to the Extended Self-Similarity. Interplay between
local and non-local interactions can be considered as a possible
hydrodynamics mechanisms of these phenomena.

\acknowledgments

I thank K.R. Sreenivasan for inspiring cooperation. Discussions with T. Nakano and V. Steinberg 
were very useful.

\end{document}